\documentclass[12pt,a4paper]{article}
\usepackage{natbib,graphicx}

\title{How do the small planetary satellites rotate?}

\author
{A. V. Melnikov\thanks{E-mail: melnikov@gao.spb.ru}\, and I. I. Shevchenko \\
Pulkovo Observatory of the Russian Academy of Sciences,\\
Pulkovskoje ave. 65/1, St.Petersburg 196140, Russia}
\begin{document}

\maketitle

\begin{abstract}
We investigate the problem of the typical rotation states of the
small planetary satellites from the viewpoint of the dynamical
stability of their rotation. We show that the majority of the
discovered satellites with unknown rotation periods cannot rotate
synchronously, because no stable synchronous 1:1 spin-orbit state
exists for them. They rotate either much faster than synchronously
(those tidally unevolved) or, what is much less probable,
chaotically (tidally evolved objects or captured slow rotators).
\end{abstract}


\bigskip

The majority of planetary satellites with known rotation states
rotates synchronously (like the Moon, facing one side towards a
planet), i.e., they move in synchronous spin-orbit resonance 1:1.
The data of the NASA reference guide \citep{NASA} combined with
additional data \citep{Maris01,Maris07,Grav03} implies that, of
the 33 satellites with known rotation periods, 25 rotate
synchronously.

For the tidally evolved satellites, this observational fact is
theoretically expected. The planar rotation (i.e., the rotation
with the spin axis orthogonal to the orbital plane) in synchronous
1:1 resonance with the orbital motion is the most likely final
mode of the long-term tidal evolution of the rotational motion of
planetary satellites \citep{GP66,P77}. In this final mode, the
rotational axis of a satellite coincides with the axis of the
maximum moment of inertia of the satellite and is orthogonal to
the orbital plane.

Another qualitative kind of rotation known from observations is
fast regular rotation. There are seven satellites that are known
to rotate so \citep{Maris01,Maris07,Grav03,Bauer04,NASA}:
Himalia~(J6), Elara~(J7), Phoebe~(S9), Caliban (U16),
Sycorax~(U17), Prospero~(U18), and Nereid~(N2); all of them are
irregular satellites. These satellites, apparently, are tidally
unevolved.

A third observationally discovered qualitative kind of rotation is
chaotic tumbling. \citet{WPM84} and \citet{W87} demonstrated
theoretically that a planetary satellite of irregular shape in an
elliptic orbit could rotate in a chaotic, unpredictable way. They
found that a unique (at that time) probable candidate for the
chaotic rotation, due to a pronounced shape asymmetry and
significant orbital eccentricity, was Hyperion~(S7). Besides, it
has a small enough theoretical timescale of tidal deceleration of
rotation from a primordial rotation state. Later on, a direct
modelling of its observed light curves \citep{K89b,BNT95,D02}
confirmed the chaotic character of Hyperion's rotation. Recent
direct imaging from the {\it CASSINI} spacecraft supports these
conclusions~\citep{T07}.

It was found in a theoretical research~\citep{KS05} that two other
Saturnian satellites, Pro\-me\-theus~(S16) and Pandora~(S17),
could also rotate chaotically (see also \citet{MS08}). Contrary to
the case of Hyperion, possible chaos in rotation of these two
satellites is due to fine-tuning of the dynamical and physical
parameters rather than to a large extent of a chaotic zone in the
rotational phase space.

We see that the satellites spinning fast or tumbling chaotically
are a definite minority among the satellites with known rotation
states. However, the observed dominance of synchronous behaviour
might be a selection effect, exaggerating the abundance of the
mode typical for big satellites. This is most probable.
\citet{P77} showed on the basis of tidal despinning timescale
arguments that the majority of the irregular satellites are
expected to reside close to their initial (fast) rotation states.

A lot of new satellites has been discovered during last years. Now
the total number of satellites exceeds 160 (see \citet{NASA}). The
rotation states for the majority of them are not known. In what
follows, we investigate the problem of typical rotation states
among all known satellites.

We consider the motion of a satellite with respect to its mass
centre under the following assumptions. The satellite is a
nonspherical rigid body moving in a fixed elliptic orbit about a
planet. We consider the planet to be a fixed gravitating point.
The shape of the satellite is described by a triaxial ellipsoid
with the principal semiaxes $a > b > c$ and the corresponding
principal central moments of inertia $A < B < C$. The dynamics of
the relative motion in the planar problem (i.e., when the
satellite rotates/librates in the orbital plane) are determined by
the two parameters: $\omega_0=\sqrt{3(B-A)/C}$, characterizing the
dynamical asymmetry of the satellite, and $e$, the eccentricity of
its orbit. Under the given assumptions, the planar
rotational/librational motion of a satellite in the gravitational
field of the planet is described by the Beletsky
equation~\citep{B65}:
$$(1 + e \cos f){{\rm d}^2 \theta \over {{{\rm d} f}^2}} - 2e \sin f \,
{{\rm d} \theta \over {{\rm d} f}} + \omega_0^2 \sin{\theta}
\cos{\theta} = 2e \sin f,$$ \noindent where $f$ is the true
anomaly, $\theta$ is the angle between the axis of the minimum
principal central moment of inertia of the satellite and the
``planet~-- satellite'' radius vector.

As follows from an analysis of the Beletsky equation (see
\citet{MS00} and references therein), for a satellite in an
eccentric orbit, at definite values of the inertial parameters,
synchronous resonance can have two centres in spin-orbit phase
space; in other words, two different synchronous resonances,
stable in the planar rotation problem, can exist. Consider a
section, defined at the orbit pericentre, of the spin-orbit phase
space. At $\omega_0 = 0$, there exists a sole centre of
synchronous resonance with coordinates $\theta=0 \mbox{ mod }
\pi$, ${\rm d} \theta/{\rm d} t = 1$. If the eccentricity is
non-zero, upon increasing the value of $\omega_0$, the resonance
centre moves down the ${\rm d}\theta/{\rm d} t$ axis, and at a
definite value of $\omega_0$ (e.~g., for $e = 0.1$ this value is
$\simeq 1.26$) another synchronous resonance appears.
Following~\citep{MS00}, we call the former synchronous resonance
(emerging at zero value of $\omega_0$) the {\it alpha} mode, and
the latter one~--- the {\it beta} mode of synchronous resonance.
Upon increasing the $\omega_0$ parameter, the alpha and beta modes
coexist over some limited interval of $\omega_0$ (the extent of
this interval depends on the orbital eccentricity), and in the
section there are two distinct resonance centres situated at one
and the same value of the satellite's orientation angle. Such a
phenomenon takes place for Amalthea~(J5) \citep{MS98,MS00}. On
further increasing the $\omega_0$ parameter, at some value of
$\omega_0$ the alpha resonance disappears, i.e., it becomes
unstable in the planar problem, and only the beta resonance
remains.

The ``$\omega_0$--$e$'' stability diagram is presented in Fig.~1.
Theoretical boundaries of the zones of existence (i.e., stability
in the planar problem) of synchronous resonances are drawn in
accordance with~\citep{M01}. Regions marked by ``Ia'' and ``Ib''
are the domains of sole existence of alpha resonance, ``II'' is
the domain of sole existence of beta resonance, ``III'' is the
domain of coexistence of alpha and beta resonances, ``IV'' is the
domain of coexistence of alpha and period-doubling bifurcation
modes of alpha resonance, ``V'' is the domain of non-existence of
any 1:1 synchronous resonance, ``VI'' is the domain of sole
existence of period-doubling bifurcation modes of alpha resonance.

\begin{figure}
\begin{center}
\includegraphics[width=0.75\textwidth]{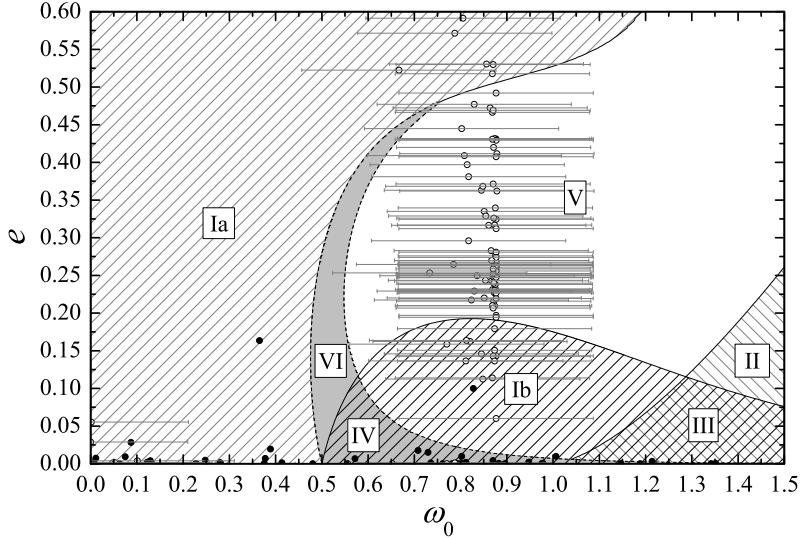}
\caption{Location of the satellites with known radii in the
``$\omega_0$--$e$'' diagram.} \label{fig1}
\end{center}
\end{figure}

The solid circles in Fig.~1 represent the satellites with known
$\omega_0$. The open circles represent the satellites with the
$\omega_0$ parameter determined by means of an approximation of
the observed dependence of $\omega_0$ on the satellite size $r$,
accomplished following an approach by \citep{MS07}. In total, the
data on sizes and orbital eccentricities are available for 145
satellites \citep{K03,SJ03,SJK05,SJK06,P07,T07,JPL}; so, there are
145 ``observational points'' in the stability diagram
``$\omega_0$--$e$'' in Fig.~1. The horizontal bars indicate
three-sigma errors in estimating $\omega_0$. They are all set to
be equal to the limiting maximum value $0.21$.

From the constructed diagram we find that 73 objects are situated
in domain V, and in domain Ib 12 objects are situated higher than
Hyperion (a sole solid circle in domain Ib), while in domain Ib
there are 15 objects in total. Synchronous state of rotation does
not exist in domain V. For the majority of satellites in domain Ib
(namely, for those that are situated higher than Hyperion)
synchronous rotation is highly probable to be attitude unstable as
in the case of Hyperion. So, 73 satellites in domain V and 12
satellites in domain Ib rotate either regularly and much faster
than synchronously (those tidally unevolved) or chaotically (those
tidally evolved). Summing up the objects, we see that a major part
(at least 85 objects) of all satellites with unknown rotation
states (132 objects), i.e., at least 64\%, cannot rotate
synchronously.

In summary, though the majority of planetary satellites with known
rotation states rotates synchronously (facing one side towards the
planet, like the Moon), a significant part (at least 64\%) of all
satellites with unknown rotation states cannot rotate
synchronously. The reason is that no stable synchronous 1:1
spin-orbit state exists for these bodies, as our analysis of the
satellites location on the ``$\omega_0$--$e$'' stability diagram
demonstrates. They rotate either regularly and much faster than
synchronously (those tidally unevolved) or chaotically (tidally
evolved objects or captured slow rotators).

With the advent of new observational tools, more and more
satellites are being discovered. Since they are all small, they
are all irregularly shaped \citep{KS06}. Besides, the newly
discovered objects typically move in strongly eccentric orbits
(see \citet{JPL,SJ03}). Therefore these new small satellites are
expected to be located mostly in domain V of the
``$\omega_0$--$e$'' stability diagram. Consequently, either fast
regular rotation (most probable) or chaotic tumbling (much less
probable), but not the ordinary synchronous 1:1 spin-orbit state,
can be a typical rotation state for the newly discovered planetary
satellites.

\end{document}